\documentclass[12pt]{article}
\input{epsf.tex}
\newlength{\dinwidth}
\newlength{\dinmargin}
\title{Study of Equilibrium Using Collision Dynamics}
\author {Amandeep Sood and Rajeev K. Puri\\
\it Physics Department, Panjab University, Chandigarh -160 014, India.\\}
\begin{document}
\maketitle
\begin{abstract}
We discuss the possibility of equilibrium (and thermalization)
 in heavy-ion collisions at
intermediate energies within a transport model. This was
 achieved by dividing
the nuclear matter into different collision zones. We find that those
 nucleons
which experience  at least ten collisions are close to complete equilibrium whereas others
never achieve any equilibrium.
\end{abstract}
\section{Introduction}
The intermediate energy heavy-ion collisions are very useful to study
the non - equilibrium dynamics of finite size systems. In addition,
one also has an  opportunity to understand the properties of hot and dense nuclear
matter that exists for a short span of time during the reaction.
It is worth mentioning that
 no direct extraction can be made  about these phenomena and
therefore,
one has to rely either on an indirect method of extraction or on a
dynamical
theoretical model that is capable of simulating
 the reaction from the start
till the end where measurements are made. The question
of equilibrium (and thermalisation) 
can be dealt within those models which are not based on the assumption of any
equilibrium.  The transport models such as Boltzmann-Uehling-Uhlenbeck
[BUU] \cite{buu} or Quantum Molecular Dynamics [QMD] \cite{qmd1,qmd2}
 are very helpful as these
models can also handle the non-equilibrated nuclear matter
formed during the early phase of the reaction. Naturally, this anisotropy in
the momentum space (that reflects the
non-equilibrium situation) should be taken into account while
 studying a heavy-ion collision.

We here plan to discuss the degree of equilibrium that can be reached
in an intermediate energy heavy ion collision.
This study is made within the framework of QMD model \cite{qmd1,qmd2}.
We shall show that the degree of
equilibrium (studied via rapidity distribution)
 depends on the reaction geometry as well as  on the number
of collisions any  individual nucleon suffers.
The preservation of the initial memory of nucleons  is directly
related with the number of collisions one suffers. Therefore,
the momentum space of those nucleons who suffer large number of collisions should be better
thermalized.

 The section 2 deals with details of the model. Section
3 depicts the results and discussion. Finally, we summarise the results in
section 4.

\section{The Model}
The nucleons in a molecular dynamics picture
interact via two- and three-body forces. The explicit two- and
three- body interactions lead to the preservation of fluctuations and
correlations that are important for N-body phenomena like
multifragmentation. This is in contrast to the one- body dynamical models
which are suitable for  one-body observable only.

In QMD model \cite{qmd1,qmd2}, each nucleon is represented by a Gaussian distribution whose
 centriod propagates with classical equations of motion:
\begin{equation}
\frac {d {\bf r}_i}{dt} = \frac{d H} {d{\bf p}_i};
\end{equation}
\begin{equation}
\frac{d{\bf p}_i}{dt} = - \frac{d H}{d{\bf r}_i},
\end{equation}

where the Hamiltonian is given by :
\begin{equation}
H = \sum_i \frac{{\bf p}_i^2}{2m_i} + V^{tot},
\end{equation}
with
\begin{equation}
V^{tot} = V^{loc} + V^{Yuk} + V^{Coul}+ V^{MDI}
\end{equation}
Here  $V^{loc}, V^{Yuk}, V^{Coul}$ and $V^{MDI}$, represent, respectively,
the Skyrme, Yukawa, Coulomb and momentum dependent (MDI) parts of the
interaction. The interaction without MDI part is called static
interaction. The different values of the compressibility
in the Skyrme force give possibility to look for the  role of different
equations of state termed as soft and hard equations of state.
 The inclusion of
momentum dependent interactions are labelled as soft momentum dependent (SMD)
and hard momentum dependent (HMD), respectively.
%%%%%%%%%%%%%%%%%%%%%%%%%%%%%%%%%%%%%%%%%%%%%%%%%%%%%%%%%%%%%%

The G- matrix at higher excitation energies becomes
complex in nature and its imaginary part acts like the collision term.
We shall use here an energy dependent nucleon-nucleon cross-section.
  It is, however,
worth mentioning that the reaction dynamics depends on the form and
magnitude of the  nucleon-nucleon cross-section \cite{qmd2}. 
%%%%%%%%%%%%%%%%%%%%%%%%%%%%%%

\section{Results and Discussion}
The present study is made by simulating the reactions of Ca-Ca,
Xe-Sn and Au-Au  at different
incident energies as well at different impact parameters. We here use a soft
equation of state along with energy dependent nucleon-nucleon
cross-section \cite{qmd1} through out the
discussion.
%%%%%%%%%%%%%%%%%%%%%%%%%

There are several different ways to define
the degree of equilibrium. The first quantity is the anisotropy ratio
$\langle{R_a}\rangle$ which is defined as \cite{ther}
\begin{equation}
\langle{R_a}\rangle = \frac{\sqrt{\langle{p_x^2}\rangle} +
\sqrt{\langle{p_y^2}\rangle}}{2\sqrt{\langle{p_z^2}\rangle}}.
\end{equation}
This anisotropy ratio is an indicator of the global equilibrium of the system.
The word {\it global} is due to the fact that this quantity does not depend on the 
local positions of nucleons and therefore, represents
the equilibrium of the whole system.
A full global equilibrium demands the anisotropy ratio to be
 close to unity.  Another way to study the
local equilibrium is to look for the 
   relative momentum of two colliding
Fermi spheres which indicates the deviation from a Fermi sphere. Note that
the  concept of local equilibrium is used by the
hydrodynamical models to simulate the heavy-ion reactions.
We shall, however, address the question of equilibrium with the
help of rapidity distribution that also  shows the
stopping of nuclear matter in heavy-ion collisions.

The rapidity distribution can be  defined as \cite{qmd1,qmd2,ther,temp}:
\begin{equation}
Y(i) = \frac{1}{2}\ln\frac {{\bf{E}}(i)-{\bf{p}}_{z}(i)}
{{\bf{E}}(i)-{\bf{p}}_{z}(i)},
\end{equation}
Where ${\bf E}(i)$ and ${\bf p_z }(i)$ are, respectively,
 the energy and the longitudinal momentum of the {\it i-th} particle.
For a full equilibrium, one should get a Gaussian shape distribution
peaked at the mid-rapidity region.  In other words, both the anisotropy ratio
and rapidity distribution are related and can give insight into the
equilibrium process of a reaction.

Using the above description,
 we plan to relate the frequency of nucleon-nucleon collisions with the degree
of equilibrium. This is achieved by dividing the rapidity distribution of 
the whole system into different collision zones. Among such zones,
 the spectator
matter (SM) consists of all those nucleons who do not suffered any
collision. We divide the participant region  into three zones : (i) the
low collision matter (LCM) which contains all nucleons who suffer less than
four collisions. (ii) the moderate collision matter (MCM) that has those
nucleons with 5-9 collisions and (iii) high collision matter (HCM) which
takes care of all those nucleons with more than ten collisions \cite{sood}.

First of all, we study the evolution of rapidity distribution.
   In figure 1, we display the evolution of rapidity
distribution dN/dY for the reaction of Xe-Sn at 400
MeV/nucleon for
different geometry.
Here we display the evolution for central (b=0 fm), semi-central
(b=4 fm) and peripheral (b=8 fm) impact
parameters.
%%%%%%%%%%%%%%%%%%%%%%%%%%%%%%%%%%
\begin{figure}[htb]
\vspace*{-1.0cm}
\centerline{\epsfysize=3.4in \epsffile{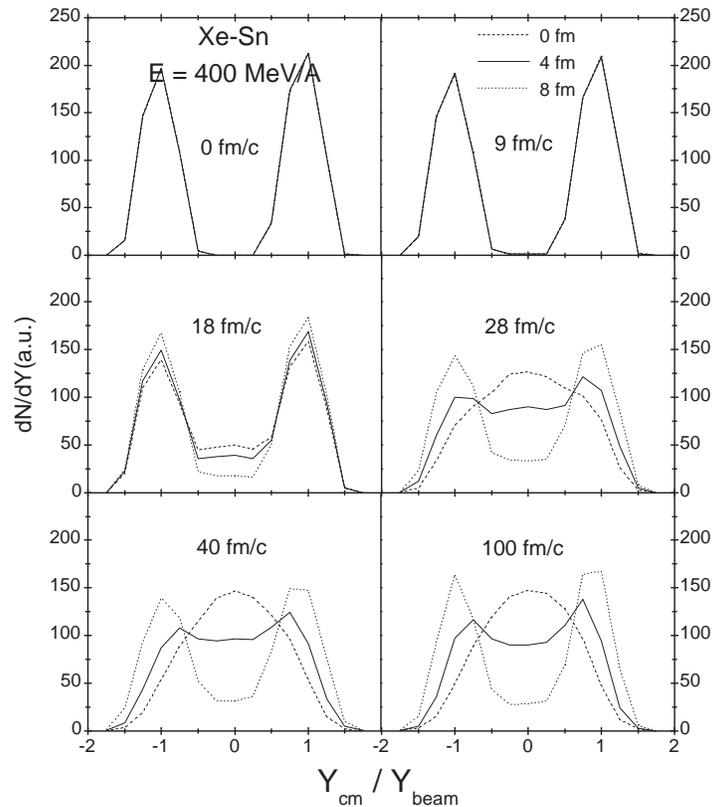}}
\vspace*{3.0cm}
\caption[]{ Rapidity distribution (dN/dY) for the reaction of Xe-Sn at
400 MeV/nucleon. We display the outcome at three impact parameters
b = 0 fm (dashed line), b= 4 fm (solid line) and b = 8 fm (dotted line).}
\label{fig1}
\end{figure}
%%%%%%%%%%%%%%%%%%%%%%%%%%%%%%%%%%%%%%%%%%

The central reaction leads to a very high nucleon-nucleon
collision rate and density
whereas the peripheral reaction has a very small overlap. At the beginning,
whole matter is grouped into either  projectile matter or into
target matter. No nucleon-nucleon collisions occur till 9 fm/c , therefore,
no change in the rapidity structure appears.
Due to some nucleon-nucleon collision between 9 and 18 fm/c, few nucleons shifts
to the mid-rapidity zone. On the other hand, drastic changes can be seen
(for the central and semi-central reactions) between 18 and 28 fm/c. 
It is that time when the density and collision rate (not shown here)
is  maximal.
The low frequency of the nucleon-nucleon collisions in
peripheral geometry does not allow substantial changes in the shape of
the rapidity distribution even at the end of the reaction.
This also points towards
the lack of degree of equilibrium in peripheral collisions.  On the
other hand, a nearly equilibrium can be seen in central collisions suggesting
better equilibrated nuclear matter in these reactions.
The study of anisotropy ratio
(not shown here) also depicts similar picture.
Its final values are close to 0.87, 0.67
and 0.37, respectively, for central, semicentral and peripheral reactions.
 As mentioned in the beginning, for a complete equilibrium, this ratio
should be equal to one. Both the rapidity distribution and the
anisotropy ratio suggest that central reactions should be better thermalised.
It is worth mentioning that both these quantities depends on the
mass of the system \cite{ther}. 

It will be of further interest to analyse the momentum space of
 nucleons of  equilibrated matter. This is achieved
by dividing the final state rapidity distribution
into the above mentioned collision zones (see figure 2).
%%%%%%%%%%%%%%%%%%%%%%%%%%%%%%%%%%%%%
\begin{figure}[htb]
\vspace*{-1.0cm}
\centerline{\epsfxsize=7.0in\epsffile{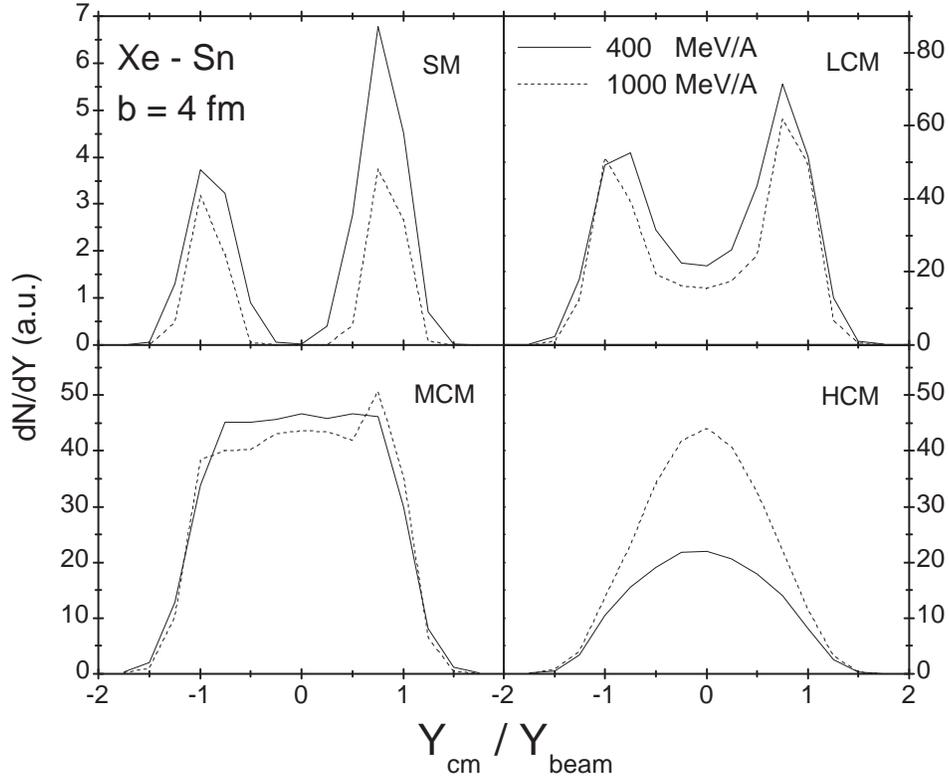}}
\vspace*{3.0cm}
\caption[]{The final state rapidity distribution dN/dY for the reaction of Xe-Sn at
400 MeV/nucleon ( solid line) and 1000 MeV/nucleon ( dashed line).  Different graphs shows
the break up of rapidity distribution into SM, LCM, MCM and HCM
zones. For details, see the text.}
\label{fig2}
\end{figure}
%%%%%%%%%%%%%%%%%%%%%%%%%%%%%%%%%%%%%%%%

Naturally, the frequency of the nucleon-
nucleon collisions increases with the incident energy, therefore,
HCM  percentage is much more at 1 GeV/nucleon than at 400 MeV/nucleon.
The maximal collision rate for Xe-Sn
reaction at 400 MeV/nucleon is 34 whereas it is 56 at 1 GeV/nucleon.
We find that the gaussian shape is more prominent for those nucleons
that suffer at least ten collisions. This is true at both incident
energies. The nucleons with less than 10 collisions observe a
partial equilibrium.
%%%%%%%%%%%%%%%%%%%%%%%%%%%%%%%%%%%%%%%%%%%%%%%%%%%%%%%%%%%%%%%
\begin{figure}[htb]
\vspace*{-1.0cm}
\centerline{\epsfxsize=7.0in\epsffile{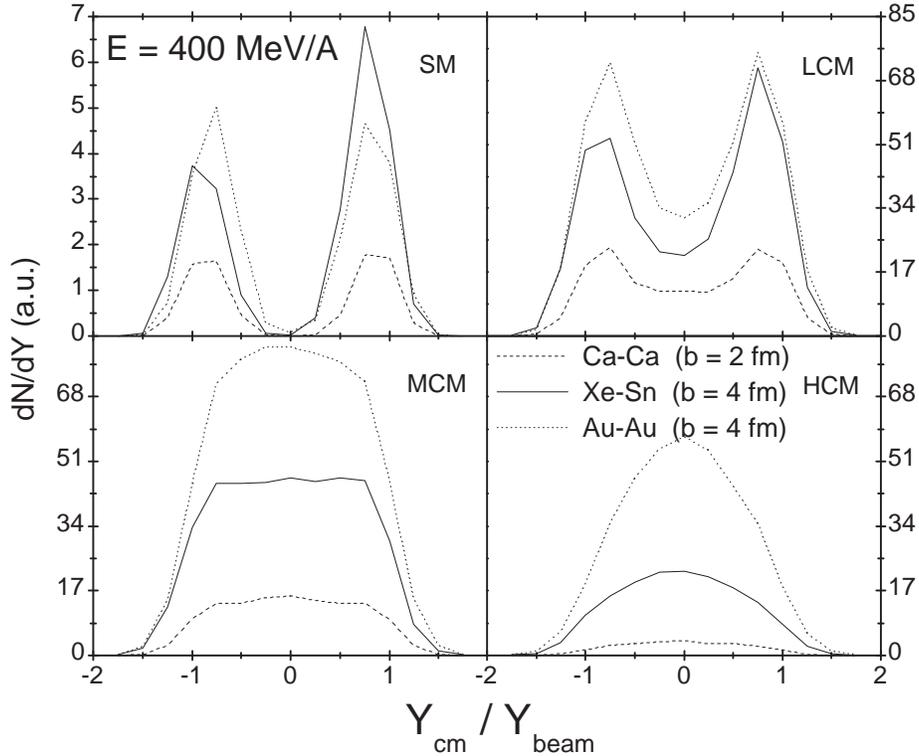}}
\vspace*{3.0cm}
\caption[]{Same as figure 2, but for the semi-central reactions
of Ca-Ca, Xe-Sn and Au-Au at 400 MeV/nucleon.}
\label{fig3}
\end{figure}
%%%%%%%%%%%%%%%%%%%%%%%%%%%%%%%%%%%%%%%%%%%%%%%%%%%%%%%%%%%%%%%%

In other words,  a complete equilibrium is possible only for those
nucleons that suffer at least 10 collisions in a reaction. The nucleon-
nucleon collisions destroy the initial correlations, therefore,  are
dominant mode of achieving the thermalization in a reaction. From figure 2,
it is also evident that the momentum distribution of nucleons with less than
4 collisions is still like that of projectile and target. 

The  mass dependence in heavy ion reactions has been found to affect the
dynamics ranging from the fusion (at low incident energy) to the multifragmentation as well as collective
flow and particle production (at intermediate and relativistsic energy) [3,7].
 To see the role of mass dependence in
thermalization, we display in figure 3, the rapidity distribution for different
collision zones using the semi-central reactions of Ca-Ca,  Xe-Sn and
Au-Au at 400 MeV/nucleon. The conclusions drawn from figure 2 are also
valid here. We find that independent of the mass of the system,
at least ten collisions are needed to form nearly equilibrated nuclear
matter. It is also evident that the heavier matter are better thermalized
compared  to lighter systems. Similar trends can also be seen from the
evolution of the anisotropy ratio.
In other words, the degree of equilibrium
depends on the size of the interacting system.
%%%%%%%%%%%%%%%%%%%%%%%%%%%%%%%%%%%%%%%%%%%%%%%%%%%%%%%%%%%%%%%%%
\begin{figure}[htb]
\vspace*{-1.0cm}
\centerline{\epsfxsize=7.0in\epsffile{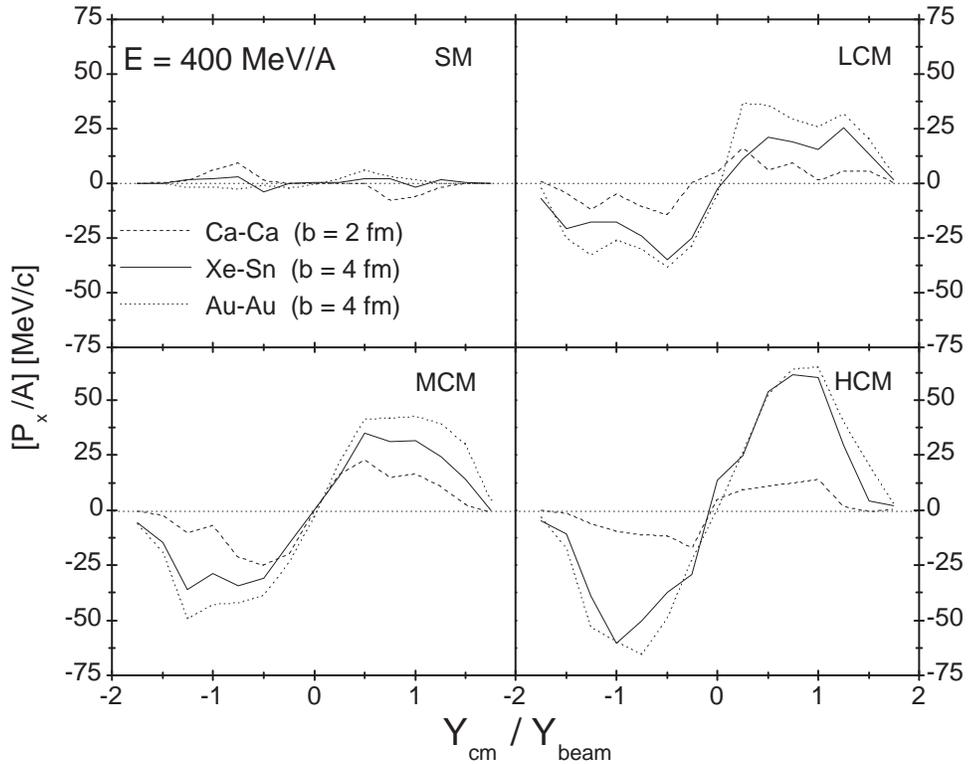}}
\vspace*{3.0cm}
\caption[]{Same as figure 3, but for $P_x$/A  at 400 MeV/nucleon.}
\label{fig4}
\end{figure}
%%%%%%%%%%%%%%%%%%%%%%%%%%%%%%%%%%%%%%%%%%%%%%%%%%%%%%%%%%%%%%%%%

In figure 4, we display the transverse momentum
$(Px/A)$ as a function of rapidity distribution for the semi-central
reactions of Ca-Ca, Xe-Sn and Au-Au at 400 MeV/nucleon. We see that the
 spectator matter does not have any transverse
momentum whereas the intensity of transverse flow increases with
collision rate. The HCM has maximum transverse flow. Further the transverse
flow depends strongly on the mass of the system which is in agreement with
all previous observations and calculations \cite{Hart}. If one extrapolates
 the above results
to very low incident energy, one will find that the balance energy ( i.e.
the energy at which flow disappears) will be lower for heavier systems
compared to lighter systems. 

\section{Conclusions}
Summarizing,
        within a dynamical quantum molecular dynamics model, we present the
study of equilibrium (i.e. thermalization) in heavy ion collisions at
intermediate energies. This was achieved by simulating the reactions
of Ca-Ca , Xe-Sn and Au-Au at different geometry as well as incident energies.
We find that the momentum space of
nucleons that suffer at least ten
collisions is better thermalization whereas nucleons with less
number of collisions exhibit a partial equilibrium. The nucleonic flow also
shows strong collision dependence.
%%%%%%%%%%%%%%%%%%%%%%%%%%%%%%%%%%%
\section*{Acknowledgement(s)}
This work is supported by the Department of Science and Tehnology, Government
of India.


\begin{thebibliography}{99}  

\bibitem{buu} G.F. Bertsch and S. Das Gupta, {\it Phys. Rep.} {\bf 160}
 (1988) 189.

\bibitem{qmd1} J. Aichelin, {\it Phys. Rep. }{\bf 202} (1991) 233.

\bibitem{qmd2} S. Kumar and R. K. Puri, {\it Phys. Rev.} {\bf C 60}
(1999) 054607; S. Kumar, R. K. Puri and J. Aichelin,
{\it Phys. Rev.} {\bf C 58} (1998) 1618;
J. Singh and R.K. Puri, Phys. Letts. {\bf B} 519, (2001) 46.

 \bibitem{ther} R.K. Puri, E. Lehmann, A. Faessler and S.W. Huang,
 {\it J. Phys. G} {\bf 20} (1994) 1817; R.K. Puri, E. Lehmann, A. Faessler
 and S.W. Huang, {\it Z. Phys. A}{\bf 351} (1995) 59.

 \bibitem{temp} R.K. Puri et al. {\it Nucl. Phys.} {\bf A 575} (1994)733.


\bibitem{sood} A. Sood and R.K. Puri, Nuclear Physics Symposium, Kolkota
(India), Dec. 26-30 (2001); A. Sood, M.Sc. Project, Panjab University, Chandigarh,
 2001 ( unpublished).

 \bibitem{Hart} Ch. Hartnack, Ph.D. thesis, University of Frankfurt (Germany),
 1993; G.D. Westfall et.al, {\it Phys. Rev. Letts.} {\bf 71}
 (1993) 1986.
\end{thebibliography}
\end{document}